# Friedel oscillations in one-dimensional metals: from Luttinger's theorem to the Luttinger liquid

Daniel Vieira, Henrique J. P. Freire, V. L. Campo Jr. and K. Capelle*

*Departamento de Física e Informática, Instituto de Física de São Carlos, Universidade de São Paulo,13560-970 São Carlos, São Paulo, Brazil*

**Abstract**

Charge density and magnetization density profiles of one-dimensional metals are investigated by two complementary many-body methods: numerically exact (Lanczos) diagonalization, and the Bethe-Ansatz local-density approximation with and without a simple self-interaction correction. Depending on the magnetization of the system, local approximations reproduce different Fourier components of the exact Friedel oscillations.
© 2007 Elsevier B.V. All rights reserved

*PACS:* 71.10.Pm; 71.10.Fd; 71.15.Mb; 75.40.Mg

*Keywords:* Friedel Oscillations; Hubbard Model; Density Functional Theory

## 1. Introduction

Any localized perturbation of a metallic system will, generically, produce density oscillations with a wave vector determined by a combination of Fermi surface effects and many-body effects. Such Friedel oscillations arise, in particular, near the boundaries of finite-size systems, and decay with a characteristic power law behaviour into the bulk.

Although density-functional theory (DFT) [1] in principle produces the correct density profile of an arbitrary inhomogeneous system, in practice DFT faces difficulties due to *(i)* the unavoidable approximations to the exchange-correlation functional, and *(ii)* the common mapping of the interacting many-body system on the auxiliary noninteracting Kohn-Sham system. In *ab initio* calculations for three-dimensional metals, e.g., the wavelength of the Friedel oscillations is determined by $2\boldsymbol{k}_F$, where $\boldsymbol{k}_F$ is the Fermi wave vector of the many-body system. The Kohn-Sham system has, ideally, the same charge and spin-density distribution as the many-body system, but need not have the same Fermi surface. To obtain the correct density oscillations from an incorrect Fermi surface one requires a highly nonlocal density functional, which removes the wrong $2\boldsymbol{k}_F^{KS}$ contribution from the Kohn-Sham Fermi surface, and replaces it by the one appropriate for the actual charge-density profile. For this reason, the wave vector of Friedel oscillations obtained by solving the Kohn-Sham equations of DFT within the local-density approximation (LDA) cannot be expected to be correct.

Interestingly, this situation changes in one dimensional systems, where the Fermi surface reduces to two points, whose distance is fixed by the total particle number alone. Hence, in a one dimensional system even a local approximation should extract, from the solution of Kohn-Sham equations, the correct oscillation frequency. Essentially, this is a consequence of Luttinger's theorem, according to which particle-particle interactions can change the shape of the Fermi body, but not its volume.

On the other hand, one-dimensional metals are not correctly described by Fermi-liquid theory, but rather fall under the universality class of Luttinger liquids, where subtle many-body effects modify the simple Fermi-liquid scenario. The challenge for DFT of one-dimensional lattice models is thus whether available approximations reproduce this intricate interplay of Fermi surface and many-body effects, and what oscillation frequencies are predicted by simple local approximations.

## 2. Friedel oscillations in the one-dimensional Hubbard model via Bethe-Ansatz local-density approximations

To investigate these questions, we turn to the one-dimensional Hubbard model, whose metallic phase obeys Luttinger-liquid phenomenology, and for which Bethe-Ansatz based local-density approximations are available, and can be compared to exact results for small systems. As local approximations we use the Bethe-Ansatz local-density approximation (BA-LDA) [2] and the Bethe-Ansatz local-spin-density approximation (BA-LSDA), both in their fully numerical version [3], and a simple Fermi-Amaldi (FA) type self-interaction correction (SIC) that can be applied on top of the local approximation.

The BA-LDA and BA-LSDA schemes have been described in Refs. [2-5]. The FA-SIC scheme simply consists in multiplying the approximate interaction energy functional $E_{int}[n] = E_H[n] + E_c^{BALDA}[n]$ by the Fermi-Amaldi correction factor [6] $1 - 1/N$, where $N$ is the total particle number, to ensure that a single fermion does not erroneously interact with itself. Here, $E_H$ is the mean-field (Hartree) energy and $E_c^{BALDA}$ is the local approximation to the correlation energy, extracted from the Bethe Ansatz. Below we describe three sets of results that are representative of a larger set of calculations, on which our conclusions are based.

These calculations are for rather small systems, with up to 15 sites, because for small systems we can diagonalize the many-body Hamiltonian exactly, by Lanczos techniques, and thus have an unequivocal benchmark for judging the performance of all aproximate schemes. Computationally, the advantage of DFT, in particular within the local-density approximation, is the ease with which it can be applied also to larger and inhomogeneous systems. In applications, this advantage must be weighted against the loss in precision due to the local approximation. Here we investigate this question for the amplitude and frequency of Friedel oscillations, by comparing to exact data.

For noninteracting finite systems, one finds analytically that boundary-induced Friedel oscillations occur with wave vector $\pi(N+1)/(L+1)$ if each site can hold up to two fermions, or wave vector $2\pi(N+1/2)/(L+1)$ if the maximum occupation is limited to one. In the thermodynamic limit ($N \to \infty$, $L \to \infty$, $n = N/L = const$) these wave vectors become $2k_F = \pi n$ and $4k_F = 2\pi n$, respectively.

For *interacting* systems in the thermodynamical limite, with theoretical maximum occupancy equal to two, Luttinger-liquid theory predicts that the oscillations occur with wave vector $2k_F$ if the interactions are weak, but $4k_F$ if they are strong enough to suppress double occupancy. Numerically, it was found that the crossover takes place around $U = 4t$ [7,8]. Here, we focus attention on this crossover region in a *finite interacting* system, where for convenience we continue to label the two Fourier components as $2k_F$ and $4k_F$, although their ratio in finite systems is not exactly 2.

Figure 1 shows a typical charge-density profile. The exact $2k_F$ oscillations are reproduced qualitatively, and with few-percent accuracy also quantitatively, by the BA-LDA. In accordance with our discussion in the introduction, Luttinger's theorem thus protects this aspect of the density profile from the otherwise unavoidable errors of local approximations.

The $4k_F$ component, on the other hand, is a strong-correlation effect whose reproduction requires more than just the correct Fermi volume. Although the exact universal exchange-correlation functional of DFT doubtlessly will produce this component, too, we have been unable to reproduce it correctly with purely local approximations. As the Fourier transform shows, the intensity of the $4k_F$ component is strongly underestimated by LDA, showing that the short wavelength component of the oscillations is not properly represented by BA-LDA – although BA-LDA properly reproduces other features of the Luttinger liquid phase, such as its total energy [4], the compressibility, and the transition to the Mott insulator [6].

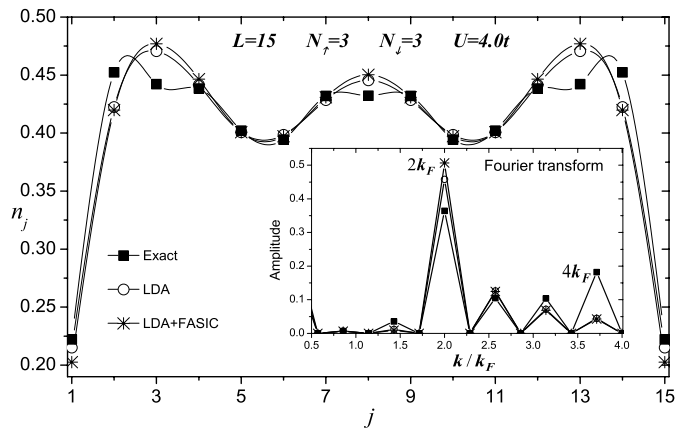

Fig. 1. Density profile of a 15-site Hubbard chain with open boundary conditions, $N_\uparrow = 3$ up spins, $N_\downarrow = 3$ down spins, and interaction $U = 4t$. Inset: Fourier transform of the density.

If a small magnetization is induced by hand in the unmagnetized system, the $4k_F$ term appears in the density profile, but at the expense of a violation of the Lieb-Mattis theorem, according to which the ground-state for even particle numbers has no net magnetization. This observation is consistent with that of Ref. [9], where it was found that in a two-site system the correct density oscillations are obtained from LSDA for contact-interacting fermions if an infinitesimal symmetry-breaking magnetic field is applied during the iterations to selfconsistency.

Figure 2 displays the charge-density profile of a system that has one more spin up particle than that of Fig. 1, so that a net magnetization $M = N_\uparrow - N_\downarrow = 1$ appears spontaneously, without violating the Lieb-Mattis theorem. The $2k_F$ contribution of the full density is now resolved in two $4k_F^\sigma$ oscillations, each arising from one spin component $n_\sigma$, but the $4k_F$ term does not split. Unlike LDA in the unmagnetized case, the LSDA now reproduces both the long ($2k_F$) and the short ($4k_F$) wavelength components of the oscillations.

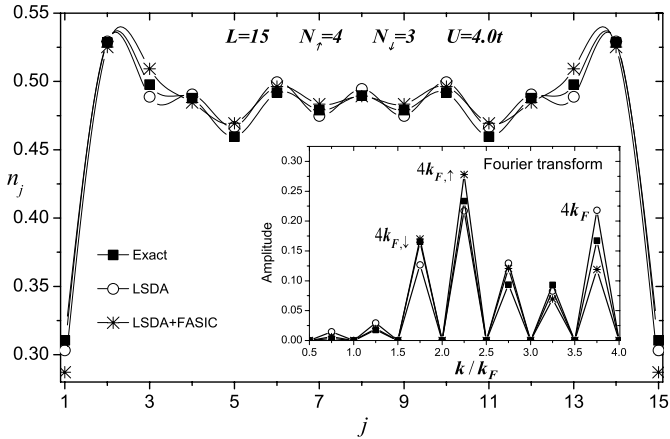

Fig. 2. Density profile of a 15-site Hubbard chain with open boundary conditions, $N_\uparrow = 4$ up spins, $N_\downarrow = 3$ down spins, and interaction $U = 4t$. Inset: Fourier transform. Differently from Fig. 1, this system has a net magnetization $M = N_\uparrow - N_\downarrow = 1$.

Figure 3 displays the local magnetization of the system of Fig. 2. The LSDA captures all relevant frequency components, and the amplitude of the oscillations is significantly improved by adding the self-interaction correction.

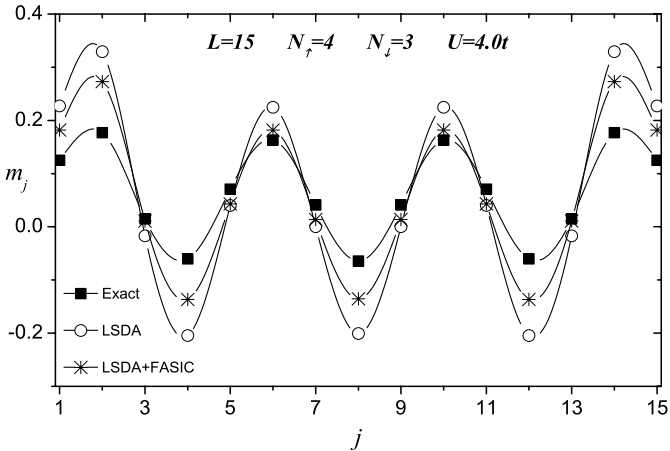

Fig. 3. Magnetization profile $m_j = n_{j\uparrow} - n_{j\downarrow}$.

## 3. Conclusions

The picture emerging from these three representative cases remains correct, in its main features, in all other situations we studied, including extension to other system sizes $L$, particle numbers $N = N_\uparrow + N_\downarrow$ and interaction strengths $U$. Local approximations have characteristic successes and failures, depending not much on these system parameters, but strongly on whether the system is magnetized or not.

Amplitudes of the charge density oscillations are qualitatively, and to within a few percent also quantitatively, correctly reproduced by BA-LDA and BA-LSDA. In the unmagnetized system, these oscillations are hardly affected by the Fermi-Amaldi self-interaction correction, whereas in the magnetized system they are sometimes slightly improved, but not consistently so. The amplitude of the magnetization density oscillations of the magnetized system, on the other hand, is consistently improved by FA-SIC.

Frequencies are harder to reproduce. In unmagnetized systems, LSDA reduces to LDA, and, due to protection by Luttinger's theorem, correctly describes the long wavelength $2k_F$ component. It fails for the short wavelength $4k_F$ component, which, however, can be obtained if the Lieb-Mattis theorem is relaxed. In magnetized systems, on the other hand, LSDA also reproduces the $4k_F$ term, without violating the Lieb-Mattis theorem. Self-interaction corrections slightly change the LDA/LSDA frequencies, but not sufficiently to recover missing Fourier components.

Future work, aiming at the complete description of Luttinger-liquid phenomenology in inhomogeneous systems, should employ nonlocal density functionals, employing gradient corrections, and/or more sophisticated self-interaction corrections. Such work is in progress.

**Acknowledgments**

This work was supported by FAPESP and CNPq. We also acknowledge computing time at the LCCA supercomputer facility of the University of São Paulo (FAPESP-2004/08928-3).